\documentclass[fleqn,usenatbib]{mnras}

\usepackage{newtxtext,newtxmath}
\usepackage[T1]{fontenc}

\DeclareRobustCommand{\VAN}[3]{#2}
\let\VANthebibliography\thebibliography
\def\thebibliography{\DeclareRobustCommand{\VAN}[3]{##3}\VANthebibliography}

\usepackage{graphicx}	% Including figure files
\usepackage{amsmath}	% Advanced maths commands
\usepackage{epsfig}
\usepackage{times}
\usepackage{color}
\usepackage{epsf}
\usepackage{natbib}
\usepackage{longtable}
\usepackage{xspace}
\usepackage{xspace}
\newcommand{\halpha}{H\ensuremath{\alpha}\xspace}
\newcommand{\hbeta}{H\ensuremath{\beta}\xspace}
\newcommand{\fitsb}{{\sc fitsb2}\xspace}
\newcommand{\kms}{\ensuremath{\mathrm{km\,s}^{-1}}\xspace}

\title[GD1400AB]{The evolutionary history of GD1400, a white dwarf-brown dwarf binary}

\author[Casewell et~al.]{S. L. Casewell$^1$\thanks{E-mail: slc25@le.ac.uk }, M.\,R. Burleigh$^1$, R. Napiwotzki$^2$, M. Zorotovic$^3$, P. Bergeron$^4$, J. R. French$^1$, J~J. Hermes$^5$, \newauthor F. Faedi$^1$,  K. L. Lawrie$^1$\\
$^1$ School of Physics and Astronomy, University of Leicester, University Rd., Leicester LE1 7RH, UK\\
$^2$ Centre for Astrophysics Research, STRI, University of 
Hertfordshire, College Lane, Hatfield AL10 9AB, UK\\
$^3$ Instituto de F\'isica y Astronom\'ia, Universidad de Valpara\'iso, Av. Gran Breta\~na 1111, Valpara\'iso, Chile\\
$^4$ D\'epartement de Physique, Universit\'e de Montr\'eal, C.P. 6128, Succ. Centre-Ville, Montr\'eal, QC, H3C 3J7, Canada\\
$^5$ Department of Astronomy \& Institute for Astrophysical Research, Boston University, 725 Commonwealth Avenue, Boston, MA 02215, USA
}

\date{Accepted XXX. Received YYY; in original form ZZZ}

\pubyear{\the\year{}}

\begin{document}
\label{firstpage}
\pagerange{\pageref{firstpage}--\pageref{lastpage}}
\maketitle

% Abstract of the paper
\begin{abstract}
GD1400AB was one of the first known white dwarf$+$brown dwarf binaries, and is the only one of these systems where the white dwarf is a ZZ Ceti pulsator. Here we present both radial velocity measurements and time series photometry, analysing both the white dwarf pulsations and the effects of irradiation on the brown dwarf. We find the brightness temperatures of  1760$\pm10$~K for the night side and 1860$\pm$10~K for the day side indicate the brown dwarf is hotter than spectra have previously suggested, although brightness temperatures calculated using a larger radius for the brown dwarf are consistent with previously determined spectral types. We also discuss the likely evolutionary pathway of this binary, and put its common envelope phase into context with the other known systems. 
\end{abstract}

% Select between one and six entries from the list of approved keywords.
% Don't make up new ones.

%%%%%%%%%%%%%%%%%%%%%%%%%%%%%%%%%%%%%%%%%%%%%%%%%%

%%%%%%%%%%%%%%%%% BODY OF PAPER %%%%%%%%%%%%%%%%%%
\section{Introduction}

Detached brown dwarf $+$ white dwarf systems allow the exploration of a variety of aspects of binary formation and evolution, including probing the known deficit of brown dwarf companions to main sequence stars \citep{mz04, grether06, metchev09}.  
%when compared to much larger radii
%($> 1000$\,AU; $10 - 30\%$; \citealt{gizis01}).
In detached systems (close or wide) the brown dwarfs themselves can be studied spectroscopically because they dominate the spectral energy distribution at near- to mid-infrared (IR) wavelengths \citep{farihi04, dobbie05, wd0137b, casewell18b, casewell20a,lew22, zhou22}. It should be noted that there are few observational constraints on brown dwarf evolutionary models at older ages, such as might be expected for most white dwarfs ($> 1$~Gyr; \citealt{pinfield06}) as determining the age of a field brown dwarf is challenging due to the age-mass-radius-luminosity degeneracy.

The closest brown dwarf $+$ white dwarf pairs provide another channel for cataclysmic variable (CV) formation \citep{politano, littlefair2, burleigh12, hernandez}, in which the substellar companion survives common envelope (CE) evolution when it is engulfed by the envelope of the red giant progenitor to the white dwarf \citep{maxted06, rappaport, parsons17}. {\it In extremis}, the closest such binaries might even represent the end state of CV evolution, in which the secondary has become highly evolved through mass transfer \citep{patterson05}.  Indeed, \cite*{zorotovic22}  calculated the CE efficiency for the known white dwarf $+$ brown dwarf pairs to be $0.24 \le \alpha_{\rm CE} \le 0.41$ which is consistent for CVs with main sequence secondary stars.

In close detached binaries, the brown dwarf is irradiated by the high ultraviolet (UV) flux of the white dwarf, leading to substantial temperature differences between the ``day'' and ``night'' hemispheres. Such systems can provide empirical laboratories for comparison with models for irradiated ``hot Jupiter'' atmospheres \citep{fortney08, knutson12, stevenson14, beatty19, arcangeli, mikal22}. However, detached brown dwarf companions to white dwarfs are rare (the fraction of L-type secondaries is $<$\,0.5\%; \citealt*{fbz05}, \citealt{girven11, steele11}) with only $\sim$20 such systems known to date, of which $\sim$10 are close, post CE binaries, although many candidates are known \citep{brown}. The majority of these systems studied to date have been those in which the brown dwarf is highly irradiated: WD0137-349B \citep{casewell15, longstaff17, lee20, zhou22}, SDSS1411 \citep{casewell18b, lew22}, EPIC212235321 \citep{casewell18, lothringer20, zhou22}. These three systems have periods between 68~min and $\sim$2~hrs, and the white dwarfs have effective temperatures between 25,000~K and 13,000~K.  Very little is known as to the effects of irradiation  on brown dwarfs orbiting cooler white dwarf primaries, although what is known suggests some form of inflation of the brown dwarf is likely (e.g. \citealt{casewell20a, casewell20b}).The majority of these close binaries have periods of $\sim$2~hrs (e.g. \citealt{maxted06, steele13, littlefair14, casewell20a}), making those with longer periods: GD\,1400AB at $\sim$10~hrs \citep{burleigh11} and ZTFJ0038+2030 with a period of 10.36~hrs \citep{ztf21} unusual. The only post-CE systems known with substellar companions and long orbital periods are Gaia~0007-1605 which has a period of 1.0446 days \citep{rebassa22} and the white dwarf-planet system WD1856+534Ab which has a period of 1.407 days \citep{vanderburg20}.

GD\,1400 (WD0145-221) is a DA white dwarf located at 46.25 $\pm$ 0.07 pc.
\citet{farihi04} determined GD\,1400 had a likely unresolved L dwarf companion, confirmed as an L7 dwarf with near-IR spectroscopy by \citet{dobbie05}. \citet{farihi05} subsequently obtained $Spitzer$ IRAC photometry from 3.6 to 9 microns which was also consistent with a secondary of L5--L7 spectral type.  The most recent work on the system was by \citet{walters}, who determined an effective temperature of 2100~K for the brown dwarf, significantly hotter than suggested by the previous spectra. GD\,1400 is known to be  a ZZ Ceti pulsator, hence it is photometrically variable with \citet{fontaine03, kilkenny14, TessCV} detecting numerous periods.

In this paper we present radial velocity measurements and time series photometry of the GD\,1400 system in order to better constrain the mass of the brown dwarf, likely levels of irradiation, and the effects the pulsations are having on the system. We also present an in-depth analysis of the likely CE evolution of the system, and how this compares to other binaries formed via the same evolutionary pathway.

\section{GD\,1400A}

\citet{gentilefusillo2021} fitted the $Gaia$ photometry and parallax of GD\,1400A with pure hydrogen atmosphere models and obtained an effective temperature of $T_{\rm eff}=11090\pm160$~K and a surface gravity of $\log g = 8.00\pm0.03$, resulting in a mass of 0.604$\pm$0.016~M$_{\odot}$. This is broadly consistent with the values given in \cite{koester09} of $T_{\rm eff}=11747\pm20$~K and $\log g = 8.066\pm0.007$ from spectroscopy alone. A more recent spectroscopic determination by \citet{bergeron21} yields a slightly higher mass with parameters of $T_{\rm eff}=11390\pm260$~K, $\log g = 8.17\pm0.19$ and a mass of 0.71 $\pm$0.06~M$_{\odot}$, although the spectrum in this case is a classification spectrum from the Montreal-Cambridge-Tololo (MCT) survey. \citet{walters} determined parameters of $T_{\rm eff}=11 000\pm500$~K and a mass of 0.59 $\pm$ 0.07~M$_{\odot}$ based on the $Gaia$ DR3 parallax and the Pan-STARRS photometry. 

\begin{figure}
\begin{center}
\includegraphics[scale=0.5,trim={2cm 17.5cm 0cm 0cm}, clip]{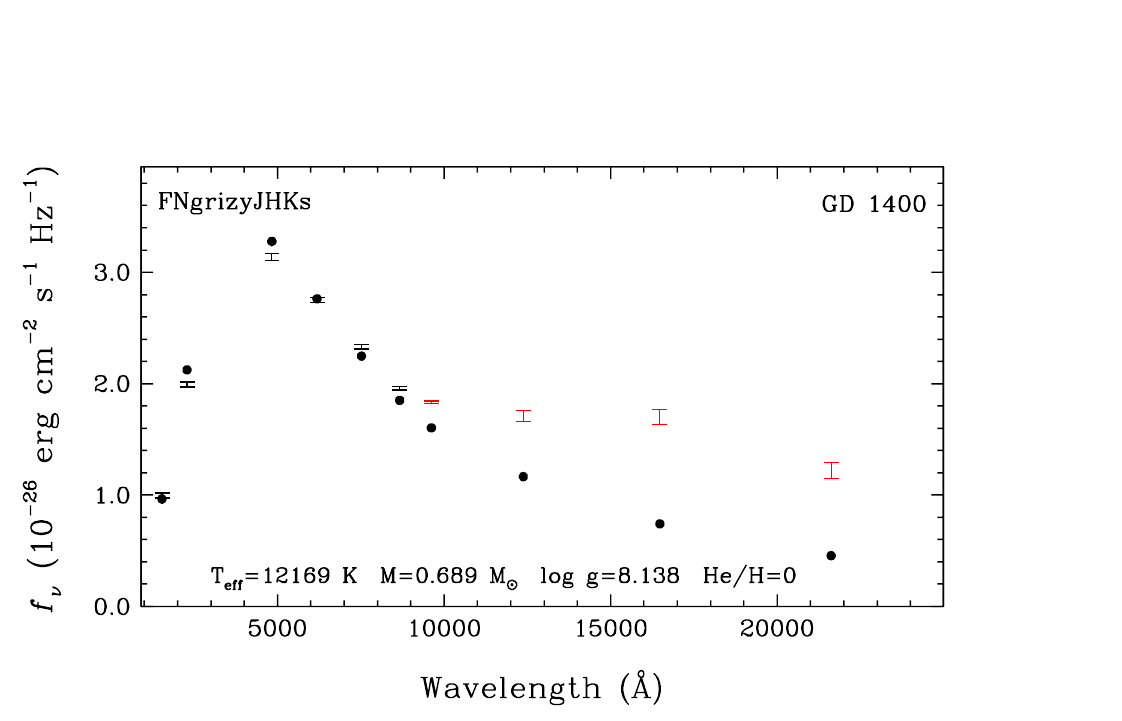}
\caption{Our best fit to the photometry from $Galex$, Pan-STARRS, and 2MASS (shown by error bars), and using the $Gaia$ DR3 parallax. Filled circles represent the best fit model with parameters given in the panel. The red error bars correspond to the photometric data not included in the fit as they are in photometric excess over the single white dwarf model.}
\label{photom}
\end{center}
\end{figure}

\citet[][see their Figure 4]{vincent} show the difference between spectroscopically and photometrically derived parameters for ZZ Ceti stars, highlighting the fact that time averaged photometry of ZZ Ceti pulsators is not the same as for non-pulsating DA white dwarfs. The lack of a $u$ band measurement for GD1400A would also likely bias the photometric fit towards lower effective temperatures \citep{bergeron19}. We decided to independently measure the parameters of GD\,1400A by reanalyzing the best available photometric and spectroscopic data available. We show in Figure \ref{photom} our photometric fit to GD\,1400A using the $Gaia$ parallax, $Galex$, Pan-STARRS $grizy$, and 2MASS photometry. In doing so, however, we exclude the photometric passbands that appear contaminated by the brown dwarf companion. We obtain $T_{\rm eff}=12169\pm50$~K, $\log g=8.14\pm0.06$, and $M=0.689^{+0.006}_{-0.004}$~M$_{\odot}$.  

\begin{figure}
\begin{center}
\includegraphics[scale=0.5,trim={4cm 4cm 4cm 7cm}, clip]{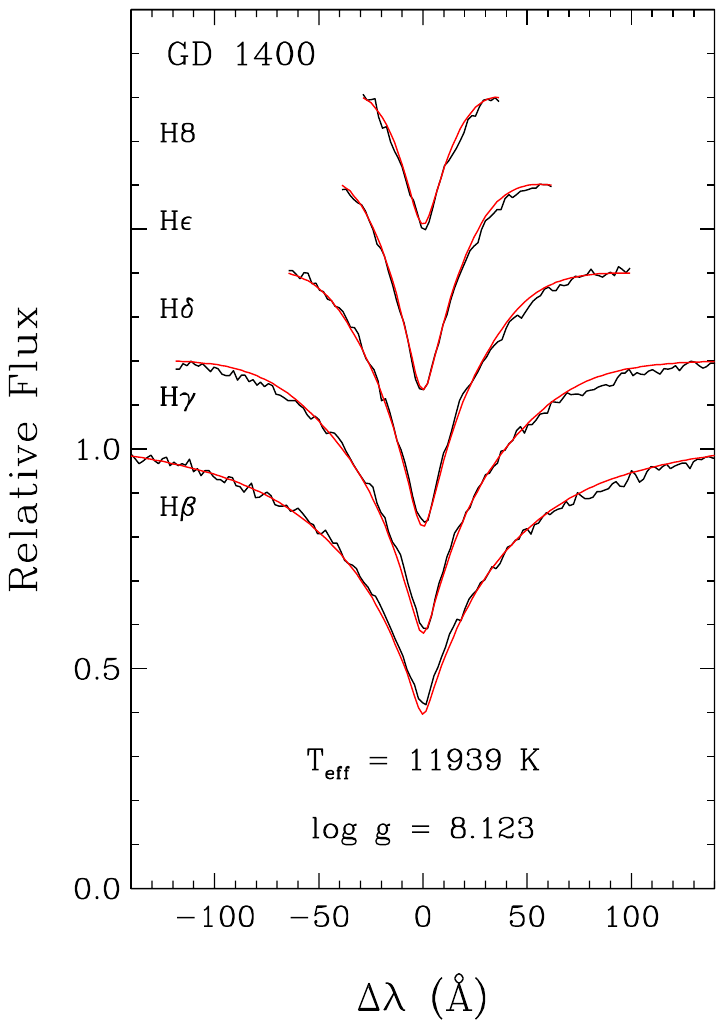}
\caption{Spectroscopic fit to the normalized Balmer lines of GD\,1400A.}\label{spec}
\end{center}
\end{figure}

We also updated the spectroscopic fit using the optical spectrum from \citet{gianninas2011} with the most recent DA white dwarf models, and by including the 3D hydrodynamical corrections from \citet{Tremblay2013}. 
We rely on the evolutionary models from \citet{bedard20} with CO cores, $q({\rm He})\equiv \log M_{\rm He}/M_{\star}=10^{-2}$, and
$q({\rm H})=10^{-4}$, which are representative of H-atmosphere white dwarfs. Our best fit displayed in Figure \ref{spec} is obtained with T$_{\rm eff}=11939\pm177$~K, $\log g= 8.123\pm0.046$, which result in a mass of M$=0.680\pm0.029$~M$_{\odot}$, in excellent agreement with our photometric solution. This updated mass gives a gravitational redshift of 35~kms$^{-1}$, larger than the measured difference in $\gamma$ velocities given in \citet{walters}, but broadly consistent with the gravitational redshift for their adopted mass. 

We can estimate an upper limit on the total age of the system using the white dwarf cooling age, the initial-final mass relation (IFMR) for white dwarfs and an estimate of the main sequence lifetime, neglecting any accelerated evolution during the CE phase. Using the software \textsc{wdwarfdate} \citep{kiman} and with the \citet{cummings} IFMR and MIST isochrones we determine the cooling age to be 0.46$^{+0.04}_{-0.03}$~Gyr and the white dwarf progenitor mass to be 2.09$^{+0.49}_{-0.52}$~M$_{\odot}$. The total system age is estimated to be 1.76$^{+1.20}_{-0.56}$~ Gyr. It should, however, be noted that this is an upper limit, as we do not know when the main sequence lifetime of the white dwarf progenitor was truncated by the common envelope evolution.

\section{Radial velocity observations}

Fifteen high resolution optical spectra of GD\,1400 were obtained between 2006 July and September with the UVES echelle spectrometer \citep{uves} on UT2 of the European Southern Observatory's Very Large Telescope (ESO VLT), under programme 077.D-0673(A). UVES was used with the DIC-1 dichroic, with the split beams centred at 3900~{\AA} and 5640~{\AA}, giving a resolution of 0.04~{\AA} and a radial velocity accuracy of 1.5\,km\,s$^{-1}$ in the cores of the H$\alpha$ and H$\beta$ absorption lines. The observations were performed in service mode, in seeing no worse than $1.4\arcsec$, for a total exposure time of 1200\,s to deliver an anticipated S/N $\approx25$ per pixel (using $2 \times 2$ binning). Each observation was split into $2 \times 600$\,s exposures to avoid smearing.

The spectra were reduced with the ESO MIDAS pipeline for UVES, in the same way as for the ESO Supernova Type 1 Survey (SPY: \citealt{napiwotzki20}) including the merging of the echelle orders and the wavelength calibration. The quality of these automatically extracted spectra is very good, except for a quasi-periodic wave-like pattern that occurs in some of the spectra. This is largely removed by additional processing. In addition, a featureless (DC) white dwarf, WD\,0000$-$345 was observed as part of the programme to aid in correcting the detector response curve.

We see no sign of emission within the Balmer features at any phase which could be caused by irradiation (e.g. \citealt{longstaff17}), which was expected as no such emission is seen in the lines of WD1032+011 \citep{casewell20a} which has a similar white dwarf temperature, but a 2~hr period.

We measured the radial velocities (RVs) of GD\,1400 from the non-local thermodynamic equilibrium (NLTE) line cores of the Balmer lines \halpha and \hbeta. 
The measurements were carried out with the package \fitsb designed to fit the spectra of single-lined (SB1) and double-lined (SB2) binaries \citep{N.Y.N2004} as for \citet{napiwotzki20}.

Synthetic spectra from \citet{koester10} were convolved to the observational resolution with a Gaussian and interpolated to the actual parameters with bi-cubic splines and interpolated to the observed wavelength scale.  We used the model profiles computed by \citet{K.D.W1998} for their investigation of rotation in white dwarfs.  These are computed performing a NLTE line formation on top of up-to-date LTE model atmospheres for DA white dwarfs \citep{K.D.W1998} for details. The inner core of GD\,1400 is not very well reproduced by the model profiles. The very likely explanation of this is GD\,1400 being a large amplitude ZZ~Ceti variable. \fitsb offers the option to combine model spectra with other line profiles. In our case we achieved reasonable representation of the observed line profile by adding a central Gaussian component.
 
The line profiles were derived from a simultaneous fit of all spectra available, significantly reducing the overall uncertainties. The RV errors were determined by bootstrapping the pixels of the spectra (see \citealt{napiwotzki20}). 
We determined a RV dispersion of $\sigma_{\mathrm{RV}}=0.71~\kms$, indicating very good stability for the observations. Any RV shifts seen within the measurements are likely caused by the ZZ Ceti pulsations \citep{berger05}.

The period search was carried out by means of a periodogram method \citep[see][]{L.M.D1998,N.E.H2001}. Sine-shaped RV curves were fitted to the measured RVs for a large range of trial periods. The quality of each fit was determined based on the sum of the squared residuals ($\chi^2$). A detailed fit for the best period estimate was then done to derive the orbital semi-amplitude $K_1$, system velocity $\gamma_1$, and the epoch of phase zero $T_0$. The latter is defined here as the conjunction time when the visible primary moves from the blue side to the red side of the RV curve.

The orbital parameters were then refined by fitting with \fitsb using the solution from above as starting point. In this process we fit all spectra simultaneously leading to a direct determination of the orbital parameters (and line profile parameters), with the errors estimated using the bootstrapping method. However, there exists no straightforward way to include the systematic errors affecting complete spectra in this error estimate, and so we used an alternative approach: the bootstrapping is applied to the set of spectra, instead of pixels within the spectra. For each bootstrapping step, a list with a random selection of spectra is produced and a fit performed exactly the same way it is done on the original set of spectra. This is repeated several times (we used 2000 iterations) and error estimates computed from the scatter of the fitted parameters. As long as ``systematic'' errors vary in a random way between spectra -- which is likely true for most error sources, e.g. centroiding errors or not perfectly corrected flexing of the spectrograph -- these are fully taken into account.

To take into account that it is possible the correct period is very different from the best fitting period, for instance outside the formal error limits, we converted the $\chi^2$ values fitted for the trial periods into probabilities before integrating over the periods -- comparing the probability for an interval centred on the primary peak with the ``outside'' region. The integration was carried out between 0.1\,d and 30\,d extending over all plausible periods for this post-CE system.

\begin{table}
\caption{Spectroscopic orbit of GD\,1400A and adopted parameters.}\label{tab:parameters}
\begin{center}
\begin{tabular}{lll}
\hline
Parameter& Value&\\
\hline
$P$ (days)              & $0.41582\pm$0.00008\\
$T_0$ (BJD)             & 2451699.888265$\pm$ 0.001128  \\ %2455498.889 ± 0.002
$K_1$ (km\,s$^{-1}$)     & $24.08\pm0.96$  \\
$\gamma_1$ (km\,s$^{-1}$)& $42.11\pm0.59$  \\
$T_1$ (K)               & 11939$\pm$180~K \\
$M_1$ (M$_{\odot}$)         & 0.680 $\pm$ 0.029\\
$M_2$~sin~$i$ (M$_{\odot}$) & 78$\pm$6 M$_{\rm Jup}$\\  
\hline
\end{tabular}
\end{center}
\end{table}

%\begin{figure}
%\centering
%\caption{Periodogram derived from the UVES radial velocity measurements}
%\includegraphics[bb=42 110 576 655,angle=270,scale=0.4]{WD0046+051_pm.ps}
%\psfig{file=gd1400_nolog.ps,angle=270,width=8.cm}
%\end{figure}

\begin{figure}
\centering

\psfig{file=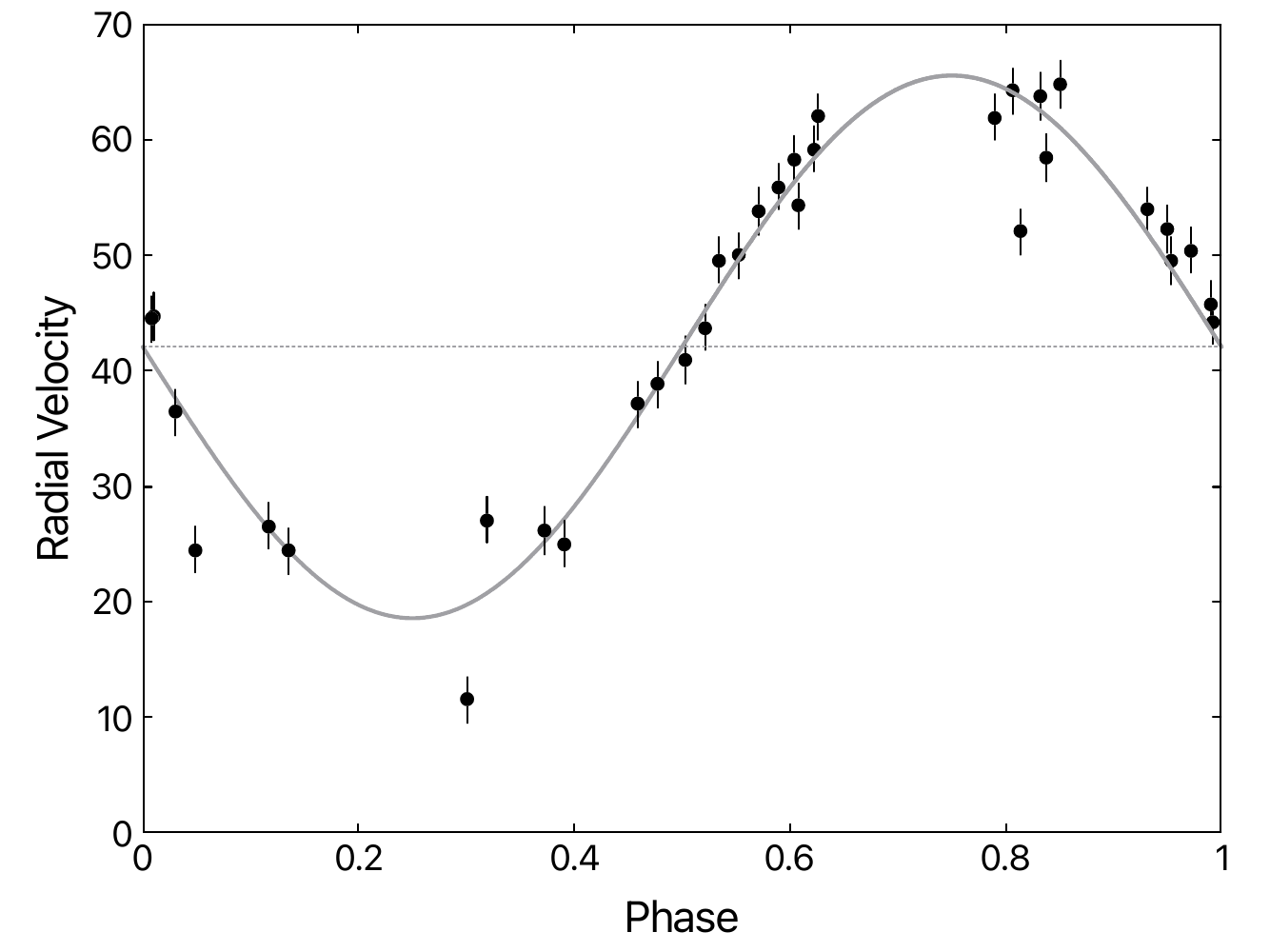,angle=0,width=8.cm}
\caption{UVES radial velocity measurements of GD\,1400 folded on the best-fit 9.98~h period.}
\end{figure}

\section{Time-series Photometry}

\subsection{Optical wavelength}
\begin{figure}
\begin{center}

\includegraphics[width=20pc]{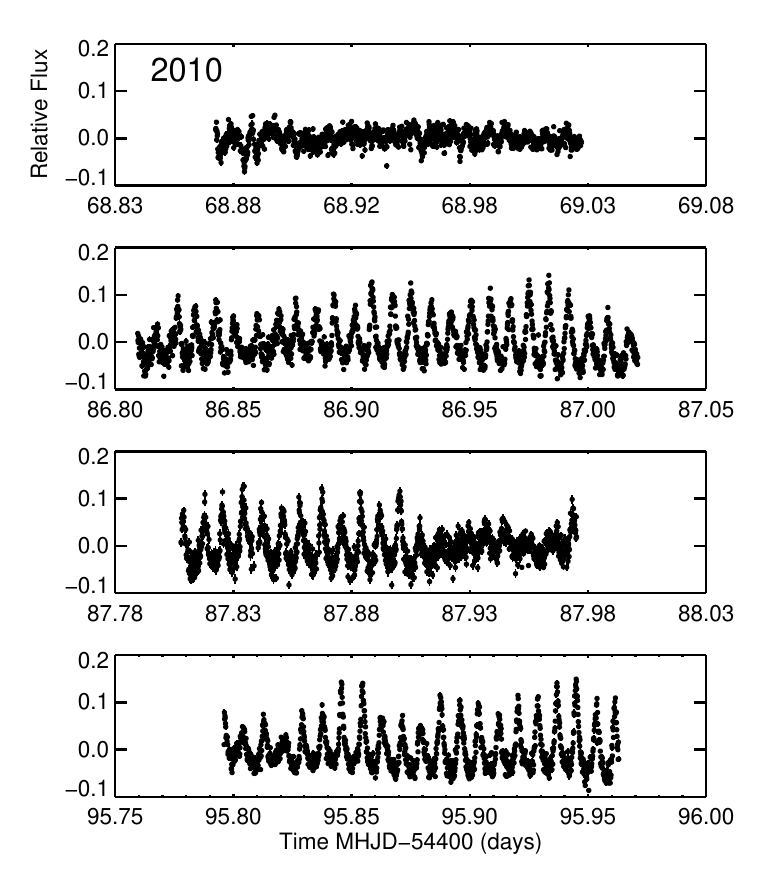}
\caption{Light curves of GD\,1400 taken in 2010 at SAAO with the UCT high-speed CCD in white light. The multiperiodic variable nature of GD\,1400 (behaviour indicative of large-amplitude ZZ Ceti white dwarfs) is clearly evident. The light curve is expressed in terms of flux relative to the star's mean brightness. Time is given in units of modified heliocentric Julian day (MHJD).}\label{fig:gd1400_2010_lc}
\end{center}
\end{figure}

\begin{figure*}
\begin{center}

\includegraphics[width=20pc]{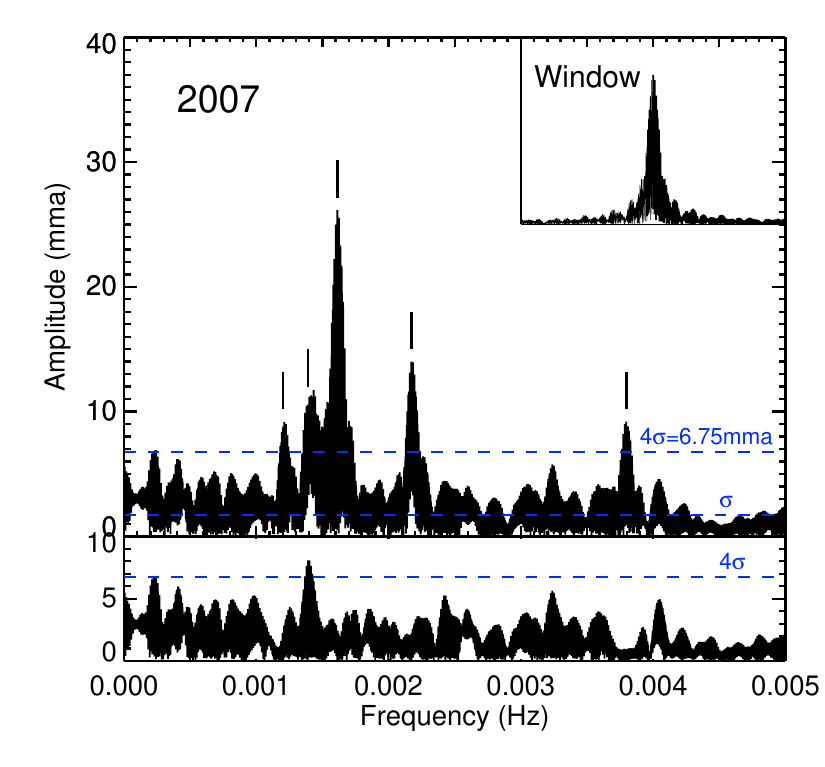}
\includegraphics[width=20pc]{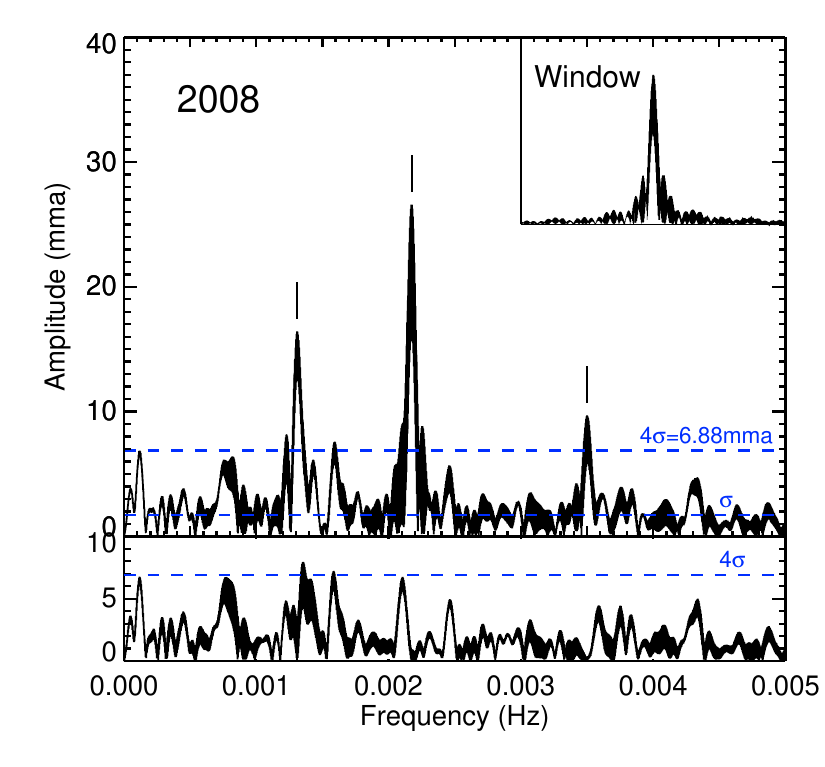}
\includegraphics[width=20pc]{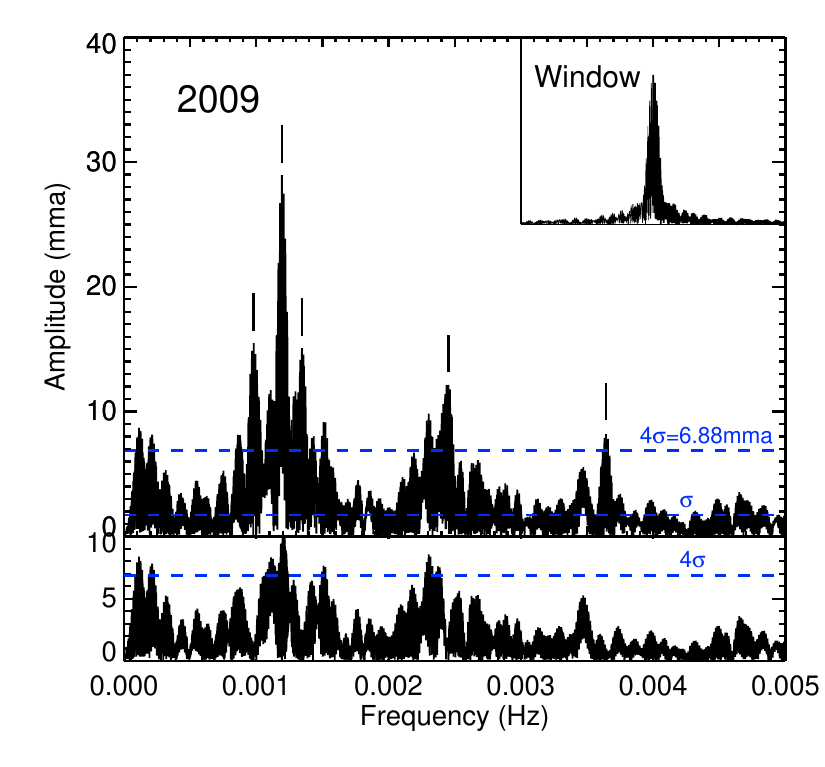}
\includegraphics[width=20pc]{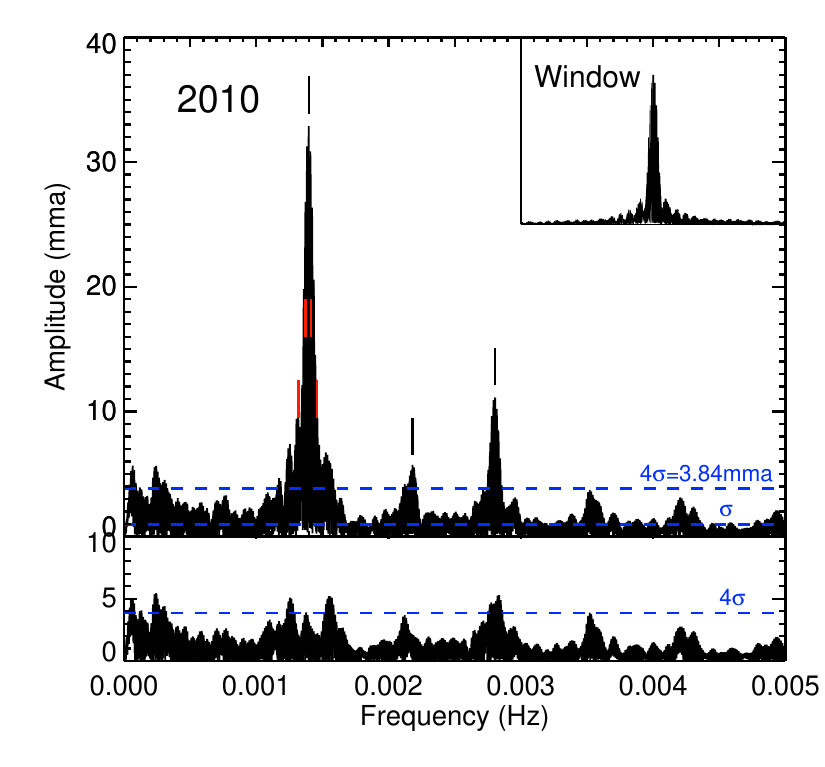}
\caption{The upper panel for each epoch of data is the Fourier Transform (FT), as calculated using Period04, where the frequencies of the pulsation modes are marked with lines. The blue, dashed lines indicate the $\sigma$ and 4$\sigma$ noise levels. The lower panel is the prewhitened FT on the same y-axis scale obtained after subtracting the indicated frequencies. The window function is given in each case. The red marks on the 2010 FT indicate multiplet best-fit periodicities. Amplitude is given in units of milli-modulation amplitude (mma), where 10\,mma is equivalent to 1\%.}\label{fig:saao_ft}
\end{center}
\end{figure*}

As  GD\,1400A is a ZZ Ceti variable, we monitored it between 2007 and 2010 at the South African Astronomical Observatory (SAAO) with the STE4 CCD imager and the UCT high-speed camera mounted on the 1.0\,m telescope in white light (i.e. with no filter). 
The data were reduced using SAAO's data reduction pipeline, which subtracted the bias, flat-fielded the science frames and extracted the brightness of the stars in the frames using {\sc DAOPHOT}. The light curve is expressed in terms of flux relative to the average brightness of the star. In some cases, a low-order, best-fit polynomial was also subtracted from the light curve to remove any residual extinction effects. A summary of the observations is given in Table \ref{tab:saao_obs} and an example of the light curves from 2010 is shown in Figure \ref{fig:gd1400_2010_lc}, where the multiperiodic nature of the variable is clearly evident. Observations taken during the same run were combined into a single epoch light curve and analysed using the Period04 program (\citealt{Lenz05}). The Fourier Transform (FT) of each epoch is illustrated in Figure \ref{fig:saao_ft}, along with its window function and prewhitened FT, obtained from subtracting the best-fit periodicities from the light curve and recalculating the FT. Table \ref{tab:saao_f_modes} summarises the best-fit periodicities determined from non-linear least-squares fitting for the 2007, 2008, 2009 and 2010 datasets, with both least squares and Monte Carlo uncertainties computed using Period04. We have not attempted to identify multiplets or $\ell$, $k$, $m$ modes and the orbital period of GD\,1400AB at $\sim$\,10\,h is not detected in these data. All phases of the orbital period have been covered with the light curve, but we also find no indication of an eclipse when the light curve is folded on the orbital period.

Tentative periods were identified by \cite{fontaine03} at 462.20\,s, 727.90\,s and 823.20\,s in data taken in July 2002 at the 3.6\,m Canada-France-Hawaii-Telescope (CFHT) with LAPOUNE, a portable Montr\'{e}al three-channel photometer. In 2012 \citet{kilkenny14} determined periods of 437\,s, 730\,s and 454\,s in order of dominance, two of which are consistent with those determined by \citet{fontaine03}.  \citet{TessCV} used 13448 TESS data points from sector 3 in 120~s mode over 20.3 days to determine periods of 415.420~s with the second highest peaks at 451~s consistent with previous work. At lower frequencies, there were clusters of periods around 1046~s, 796~s, and 766~s detected, all of which are complex (below frequencies $\sim$1400 $\upmu$Hz). They also comment that there could be a ``triplet'' of frequencies at $\sim$ 765~s leading to a rotation period of either 3.3 or $\sim$0.57 days, although the authors were unable to confirm this.  When TESS returned to Southern hemisphere observations, it was found by \citet{Bognar23} that the detected pulsation periods had changed.

We also detect three significant frequencies, at 716.34 s (1395.9763 $\upmu$Hz), 1413.0393 $\upmu$Hz and 1370.4765 $\upmu$Hz. However, if those are from splitting caused by the rotation period, they are uneven -- 17.1 $\upmu$Hz and 25.5 $\upmu$Hz from the central component, respectively. At these periods, the mode density also increases, leading to overlapping $\ell=1$ and $\ell=2$ modes making identification of triplets challenging.

GD\,1400 sits in a typical place within the ZZ Ceti instability strip \citep{hermes17},  however, the best-fit periodicities are not stable, with the peak frequencies and amplitudes changing from year to year. Unfortunately, this makes GD\,1400A an unsuitable candidate for a long-term O -- C study to search for tertiary, lower mass companions in wide orbits through a search for periodic variations in the arrival time of these pulsations \citep{MullallyP, hermes18}.

\begin{table}
\caption{Journal of SAAO time-series photometric observations.}\label{tab:saao_obs}
\begin{center}
\begin{tabular}{ccccc}
\hline
Telescope/ & Observation & Start Time & Exposure & Number  \\
Instrument & Date (UTC)  & (UTC)      & Time (s) &  of points \\
\hline
\hline
SAAO 1.0m STE4 & 2007-11-30 & 19:50:13 & 12 & 480  \\
SAAO 1.0m STE4 & 2007-12-03 & 20:18:45 & 12 & 140  \\
SAAO 1.0m STE4 & 2007-12-04 & 21:06:52 & 12 & 287  \\
SAAO 1.0m UCT  & 2008-10-31 & 18:21:50 & 10 & 1815 \\
SAAO 1.0m UCT  & 2008-11-04 & 18:35:59 & 10 & 451  \\
SAAO 1.0m UCT  & 2009-10-30 & 22:21:46 & 10 & 1438 \\
SAAO 1.0m UCT  & 2009-10-31 & 18:31:52 & 10 & 876  \\
SAAO 1.0m UCT  & 2009-11-04 & 20:55:02 & 10 & 1106 \\
SAAO 1.0m UCT  & 2010-09-29 & 20:42:01 & 10 & 1325 \\
SAAO 1.0m UCT  & 2010-10-17 & 19:18:37 & 10 & 1817 \\
SAAO 1.0m UCT  & 2010-10-18 & 19:08:50 & 10 & 1400 \\
SAAO 1.0m UCT  & 2010-10-26 & 18:59:28 & 10 & 1437 \\
\hline
\end{tabular}
\end{center}
\end{table}

\begin{table}\label{tab:saao_f_modes}
\caption{Best-fit periodicities for each epoch of data (also marked on the FTs in Figure \ref{fig:saao_ft}). Both formal least-squares uncertainties and italicised Monte Carlo uncertainties, as computed using Period04, are given.}
\begin{center}
\begin{tabular}{clccc}
\hline
Year &           & Frequency         & Period            & Amplitude \\
     &           & ($\mu$Hz)              & (s)               & (mma)     \\
\hline
\hline
2007 & {\bf 1:} & $1613.126\pm0.032$ & $619.914\pm0.012$   & $26.14\pm0.85$   \\
     &          & $\pm$\,{\it 0.041} & $\pm$\,{\it 0.016}  & $\pm$\,{\it 0.98}\\
     & {\bf 2:} & $2172.433\pm0.058$ & $460.313\pm0.012$   & $14.58\pm0.85$   \\
     &          & $\pm$\,{\it 7.514} & $\pm$\,{\it 1.593}  & $\pm$\,{\it 1.62}\\
     & {\bf 3:} & $1389.348\pm0.177$ & $719.762\pm0.092$   & $10.53\pm0.92$   \\
     &          & $\pm$\,{\it 8.420} & $\pm$\,{\it 4.363}  & $\pm$\,{\it 3.22}\\
     & {\bf 4:} & $3797.264\pm0.092$ & $263.348\pm0.007$   & $9.24\pm0.84$    \\
     &          & $\pm$\,{\it 16.074}& $\pm$\,{\it 1.115}  & $\pm$\,{\it 1.88}\\
     & {\bf 5:} & $1203.470\pm0.099$ & $830.931\pm0.069$   & $8.74\pm0.85$    \\
     &          & $\pm$\,{\it 59.798}& $\pm$\,{\it 41.390} & $\pm$\,{\it 1.93}\\
\hline
2008 & {\bf 1:} & $2176.060\pm0.038$ & $459.546\pm0.008$   & $26.62\pm0.64$   \\
     &          & $\pm$\,{\it 0.040} & $\pm$\,{\it 0.009}  & $\pm$\,{\it 0.63}\\
     & {\bf 2:} & $1307.978\pm0.060$ & $764.539\pm0.036$   & $17.75\pm0.64$   \\
     &          & $\pm$\,{\it 37.926}& $\pm$\,{\it 22.187} & $\pm$\,{\it 2.96}\\
     & {\bf 3:} & $3500.282\pm0.103$ & $285.691\pm0.009$   & $9.81\pm0.64$    \\
     &          & $\pm$\,{\it 14.089}& $\pm$\,{\it 1.150}  & $\pm$\,{\it 1.46}\\
\hline
2009 & {\bf 1:} & $1193.327\pm0.021$ & $837.994\pm0.015$   & $28.97\pm0.69$   \\
     &          & $\pm$\,{\it 0.019} & $\pm$\,{\it 0.014}  & $\pm$\,{\it 0.91}\\
     & {\bf 2:} & $980.043\pm0.040$  & $1020.363\pm0.042$  & $14.53\pm0.68$\\
     &          & $\pm$\,{\it 6.781} & $\pm$\,{\it 7.061}  & $\pm$\,{\it 1.50}\\
     & {\bf 3:} & $1343.156\pm0.046$ & $744.515\pm0.025$   & $13.02\pm0.69$\\
     &          & $\pm$\,{\it 11.537}& $\pm$\,{\it 6.396}  & $\pm$\,{\it 2.07}\\
     & {\bf 4:} & $2453.043\pm0.047$ & $407.657\pm0.008$   & $11.54\pm0.68$\\
     &          & $\pm$\,{\it 0.049} & $\pm$\,{\it 0.009}  & $\pm$\,{\it 0.77}\\
     & {\bf 5:} & $3643.886\pm0.068$ & $274.432\pm0.005$   & $8.32\pm0.68$\\
     &          & $\pm$\,{\it 0.066} & $\pm$\,{\it 0.005}  & $\pm$\,{\it 0.64}\\
\hline
2010 & {\bf 1:} & $1395.9980\pm0.0050$ & $716.3334\pm0.0025$ & $28.76\pm0.48$\\
     &          & $\pm$\,{\it 0.0056}  & $\pm$\,{\it 0.0029} & $\pm$\,{\it 0.62}\\
     & {\bf 2:}$^{\ast}$  & $1412.9989\pm0.0068$ & $707.7141\pm0.0034$ & $15.01\pm0.48$\\
     &                  & $\pm$\,{\it 0.0111}  & $\pm$\,{\it 0.0055} & $\pm$\,{\it 0.47}\\
     & {\bf 3:}$^{\ast}$  & $1370.5606\pm0.0086$ & $729.6288\pm0.0046$ & $14.44\pm0.45$\\
     &                  & $\pm$\,{\it 0.0064}  & $\pm$\,{\it 0.0034} & $\pm$\,{\it 0.45} \\
     & {\bf 4:} & $2805.1524\pm0.0068$ & $356.4872\pm0.0009$ & $10.86\pm0.38$\\
     &          & $\pm$\,{\it 0.0075}  & $\pm$\,{\it 0.0010} & $\pm$\,{\it 0.47}\\
     & {\bf 5:}$^{\ast}$  & $1317.7424\pm0.0107$ & $758.8736\pm0.0062$ & $8.57\pm0.39$\\
     &                  & $\pm$\,{\it 0.0100}  & $\pm$\,{\it 0.0058} & $\pm$\,{\it 0.44}\\
     & {\bf 6:}$^{\ast}$  & $1457.8933\pm0.0087$ & $685.9213\pm0.0042$ & $7.89\pm0.43$\\
     &                  & $\pm$\,{\it 0.0093}  & $\pm$\,{\it 0.0045} & $\pm$\,{\it 0.51}\\
     & {\bf 7:} & $2180.7316\pm0.0135$ & $458.5617\pm0.0029$ & $5.50\pm0.38$\\
     &          & $\pm$\,{\it 0.0147}  & $\pm$\,{\it 0.0031} & $\pm$\,{\it 0.40} \\
\hline
\end{tabular}
\end{center}
$^{\ast}$ Multiplet best-fit periodicities (also marked on its FT) that were required to remove the signal from the FT to approximately $\lesssim$4$\sigma$.
\end{table}

\subsection{Near-IR wavelengths}
We observed GD\,1400 using Son OF Issac (SOFI \citealt{sofi}) on the New Technology Telescope La Silla on the nights of 25 October 2007 and 26 October 2007 as part of programme 080.C-0587(A). Photometry was obtained in the  $J$, $H$ and $K_s$ filters using a 5 point dither pattern with exposure times of 4~s for $JH$ and 9~s for $K_s$. The seeing was between 0.5 and 1.6".

The data were reduced using the \textsc{starlink} software package \textsc{orac-dr} to perform the flat fielding, sky subtraction and mosaic combining each 5 dithered frames using the method outlined in \citet{casewell15}. Object extraction was performed using aperture photometry routines within SExtractor and an aperture equivalent to the seeing.

\begin{figure}
	\includegraphics[scale=0.24]{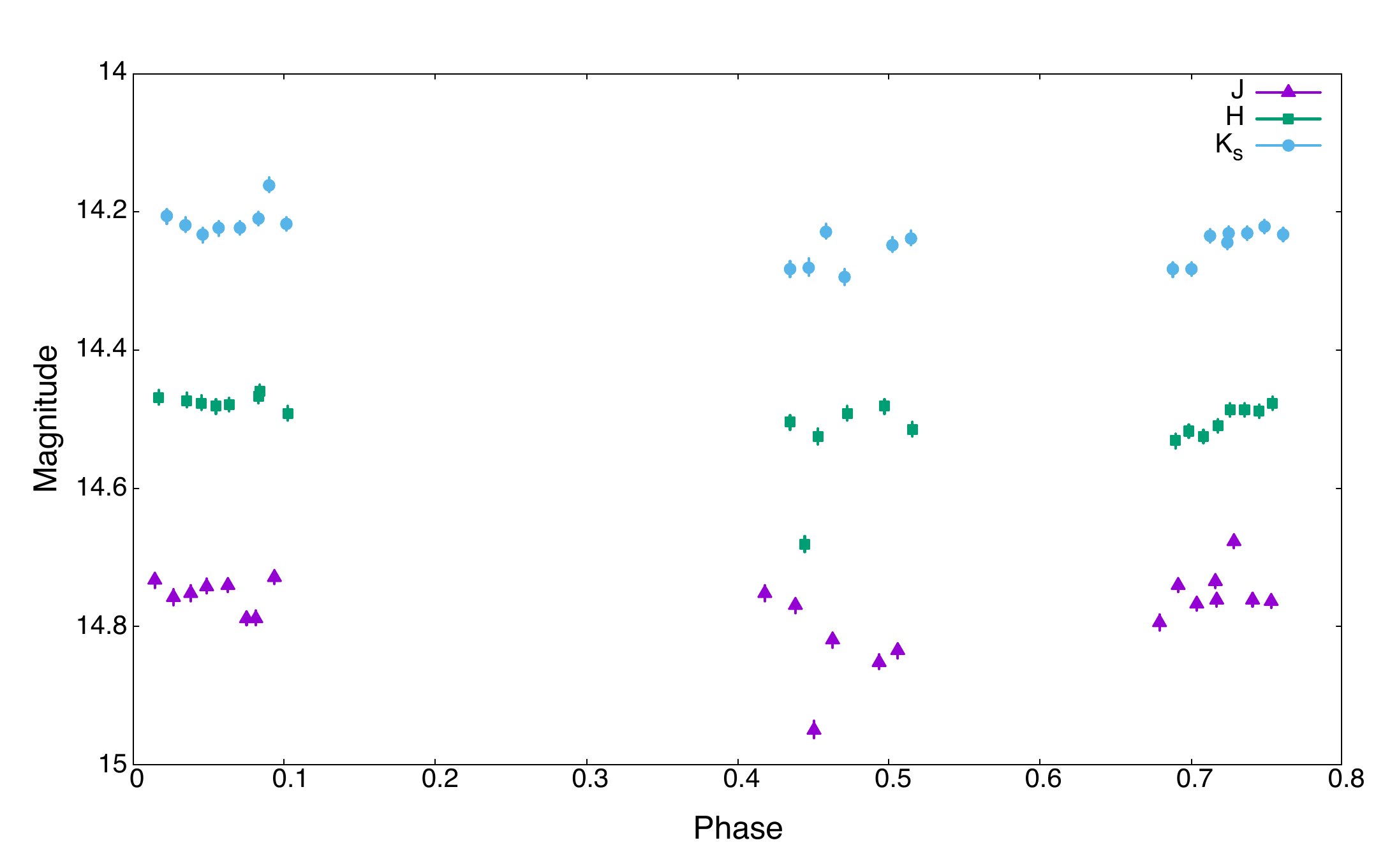}
    \caption{$J$ (purple triangles), $H$ (green boxes) and $K_s$ (blue circles) photometry phase folded on the orbital period of GD1400AB. No significant variability due to a reflection effect is detected, although the orbital sampling is poor.}
    \label{fig:nir}
\end{figure}

There are 22 $J$ band, 22 $H$ band and 23 $K_s$ band data points after reduction (Figure \ref{fig:nir}) covering phases 0-0.1, 0.4-0.55 and 0.65-0.77. The maxima and minima of the orbit are covered, however there is no reflection effect detected similar to that seen in WD0137-349AB \citep{casewell15, zhou22}, which is perhaps to be expected from the longer orbital period and cooler host star.

\subsection{Mid-IR wavelengths}

GD\,1400 has been observed by the $WISE$ mission \citep{wise, neowise} and there is archival time series photometry from  $ALLWISE$ at both W1 and W2 (3.6 and 4.5 microns). The $AllWISE$ source catalogue (combined photometry) has 13.801$\pm$0.026 and 
13.633$\pm$0.033 respectively for magnitudes in W1 and W2, broadly consistent with the 13.65$\pm$0.06 and 13.68$\pm$0.06 determined by \citet{farihi05} (it should be noted here that the $Spitzer$ and $WISE$ bands are not identical).

\begin{figure}
\centering

%\mbox{\includegraphics[angle=270,scale=0.291]{flare4.ps}}
\mbox{\includegraphics[angle=0,scale=0.4]{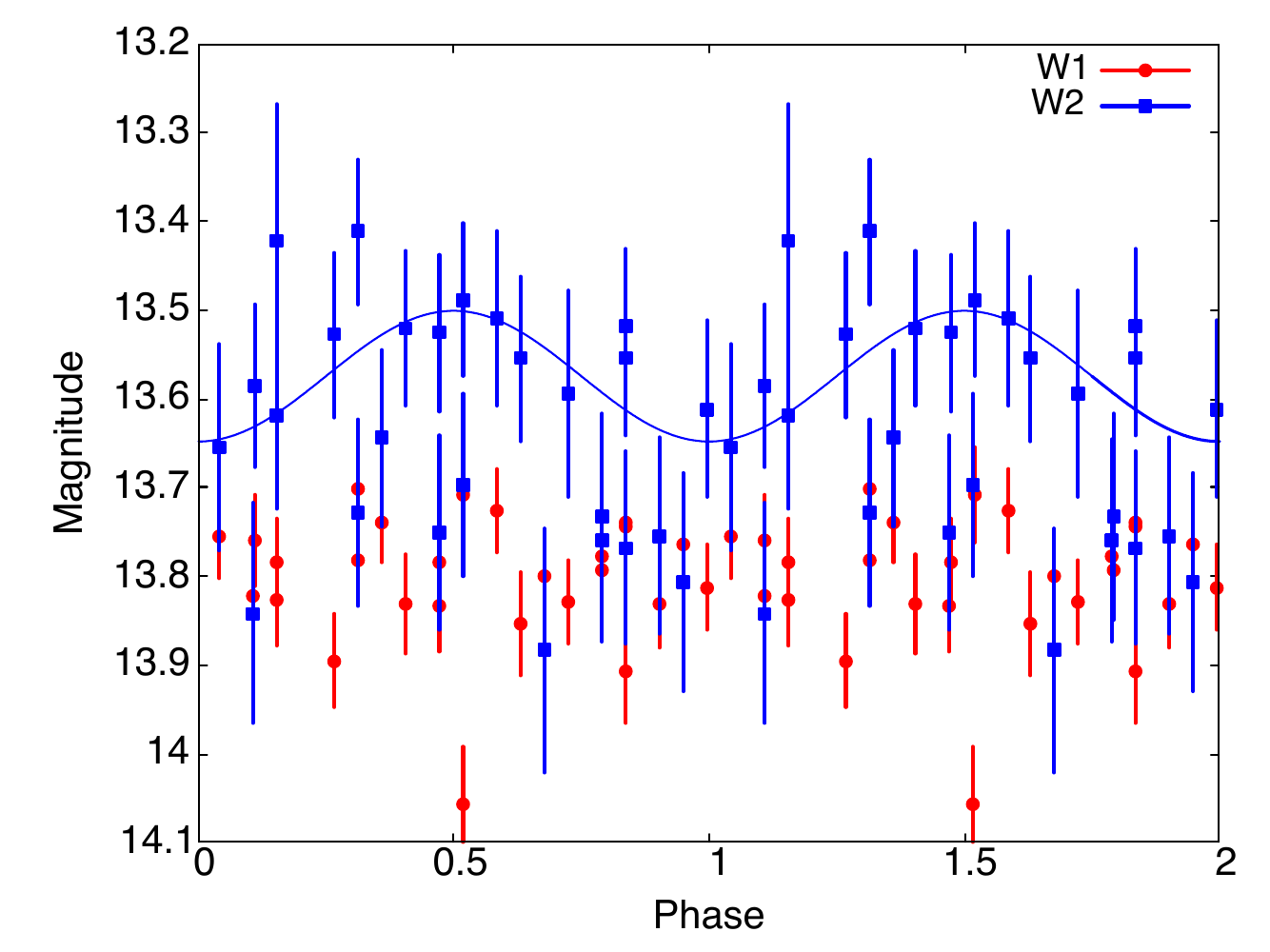}}
\caption{$AllWISE$ lightcurve for W1 (red circles) and W2 (blue boxes) phase folded on the system period showing a reflection effect in the W2 band and the best fit sine curve (blue.}\label{fig:spitzer}
\end{figure}

We used the $ALLWISE$ guidance notes and removed all photometry flagged as affected by the moon, and those flagged as poor quality, or close to the South Atlantic Anomaly. 
For each data band, we performed a sinusoidal curve fit to  the phasefolded data, keeping the period as a fixed parameter (Figure \ref{fig:spitzer}). We calculated the chi-squared metric between the best-fitting sinusoid and the data, as well as a chi-squared metric between the data and a flat line at the mean value. From this comparison we find that the reduced chi-squared value in W1 is 28.05 for a sine and 31.33 for a flat value. So W1 is 1.12 times more likely to follow a sinusoidal trend than a flat line, which is not a statistically significant result.For W2, we find the reduced chi-squared value is 9.91 for a sine and 30.79 for a flat line and that the data is 2.92 times more likely to follow a sinusoidal trend than the flat line at the mean of 13.69 mags. Therefore the W2 band shows a small reflection effect with semi-amplitude of 0.07 mags.  This reflection effect is smaller than that detected for WD0137-349B which has a semi-amplitude of 0.34 mags at 4.5 microns. NLTT5306B however, which has a 101 minute orbit, has a 4.5 micron reflection effect half that seen for GD\,1400 at 0.047 mags. GD\,1400 receives 9000 times less irradiation than WD0137-349B, and 320 times less than NLTT5306B. The reason for this difference may be clouds. As an L6-7 dwarf GD\,1400B is predicted to be more cloudy than the L5 NLTT5306B \citep{casewell20b}.

We interpolated the DA white dwarf models in the $WISE$ filters from \citet{tremblay11} for an 11,900~K log $g$=8.0 white dwarf to  obtain apparent magnitudes of W1=15.325
and W2=15.39.  We subtracted the predicted white dwarf W2 magnitude from the maximum and minimum measured by $AllWISE$ and calculated the brightness temperatures as described in \citet{casewell15} using the orbital separation of 0.009~AU and a brown dwarf radius of 0.086 R$_{\odot}$ gives brightness temperatures of 1760$\pm10$~K for the night side and 1860$\pm$10~K for the day side. If we use a larger secondary radius, such as might be expected for a 80~M$_{\rm Jup}$ object, then the temperatures drop to $\sim1550$~K on the nightside and $\sim1650$~K on the dayside. These brightness temperatures are just about consistent with those suggested by a L6-L7 dwarf.  From observations of field L~dwarfs, \citet{dupuy17} give the effective temperature of L6-L7 dwarfs as between $1441 - 1615$~K. A 1850~K dwarf would have a spectral type nearer to an L2-L3 dwarf using the same relationships.

Using the predicted magnitudes from \citet{dupuy12} for L6-7 dwarfs, we get 14.35$\pm$0.37 and 14.07$\pm$0.37 as the combined white dwarf$+$brown dwarf magnitudes of the system, just consistent with the $AllWISE$ photometry on the nightside of the brown dwarf at 4.5 microns. The measured 3.6 micron magnitudes are however too bright by $\sim$ 0.5 mags, even taking into account the large scatter on the \citet{dupuy12} magnitudes and the data. The spectral type would need to be nearer to L3-L4 to be consistent with these values.

\citet{farihi05} suggested the [3.6]-[4.5] magnitude colour of GD\,1400B is too blue when compared to a L5-L7 dwarf, suggesting that there is significant absorption in the [4.5] band due to CO, although this result does not take into account the fact that there is photometric variability at 4.5 microns. These $Spitzer$ data were taken at phases $\sim$0.15-0.18 covering a very small part of the orbit as the reflection effect is reaching its maximum. Our W1-W2 colour varies from 0.36 mags to -0.05 mags, the latter of which is consistent with the \citet{farihi05} values.

\section{Discussion}

UVES radial velocity measurements presented here reveal that the white dwarf $+$ L6-L7 brown dwarf GD\,1400AB has an orbital period $P_{\rm orb} = 0.4158$~days~$= 9.98$~hours and a separation $a=0.009$\,AU. The envelope of the post main sequence progenitor would have extended beyond this distance, so GD\,1400B must have survived a phase of CE evolution.

\subsection{The mass of the brown dwarf}

As our UVES spectra of  GD\,1400 are single lined we can only determine a lower limit on the mass of GD\,1400B from our spectroscopic data.  The lack of any features (emission or absorption) from the brown dwarf in the majority of the UVES spectra means that we can only make radial velocity measurements for the white dwarf.

Using the $K$ velocities of 21.8$\pm$1.1 and 199.2$\pm$0.6 kms$^{-1}$ presented in \citet{walters} which are consistent with our values for the white dwarf, and our updated mass for GD1400A we derived a brown dwarf mass of 0.074$\pm$0.007 M$_{\odot}$ equating to 78 $\pm$6 M$_{\rm Jup}$. 

%The total system age is estimated to be 1.76$^{+1.20}_{-0.56}$~ Gyr.
If we simply use our own $K_1$ value for the white dwarf, then from Kepler's laws and using the white dwarf mass estimate, the orbital period and our measured white dwarf's velocity $K_1 = 24.08\pm0.96$~km s$^{-1}$ ($= V_{\rm WD}$), we determine the brown dwarf mass to be $M_{\rm 2} = 0.0812\pm0.0089$ M$_{\odot}$  if we use the inclination of 60$\pm$10$^{\circ}$ \citet{walters}. At the higher end of the inclination (70$^{\circ}$), but not large enough to eclipse, the secondary mass decreases to 0.071 M$_{\odot}$.
Both these mass estimates are large for a substellar object. Indeed, the spectral type of the brown dwarf (GD\,1400B)  has been constrained by two near-IR spectroscopic observations \citep{farihi04,dobbie05} and additional {\it Spitzer} mid-IR photometry \citep{farihi05} to be L6-7. From observations of field L~dwarfs, \citet{dupuy17} give the effective temperature of L6-L7 dwarfs as between $1441 - 1615$~K. At an age of $2$~Gyr, the Sonora Bobcat \citep{Sonora} models predict the brown dwarf mass should be between 0.058 and 0.065~M$_{\odot}$. At the lowest end of our age limit this is 0.048-0.055~M$_{\odot}$, and at 6~Gyr it is 0.069-0.071~M$_{\odot}$. It should be noted the  rms scatter on the \citet{dupuy17} relationship between T$_{\rm eff}$ and spectral type is $\sim$80~K, however taking this into account, and the mass derived from the radial velocities, this would suggest an age for GD\,1400AB of greater than 3~Gyr.

If GD\,1400B is indeed only 2~Gyr old as we determine from \textsc{wdwarfdate}, then it may have had its cooling slowed by the influx of irradiation from  GD\,1400A, similar to the suggestions for NLTT5306AB \citep{casewell20b, amaro23}, or indeed be inflated such as WD1032+011B \citep{casewell20a}. This scenario would also possibly explain the mismatch between the proposed spectral type and the best fitting model as determined by the $K$ band spectra in \citet{walters}, where the model is $\sim$600~K hotter than would normally be assumed for an L6-L7 dwarf. Alternatively, both WD0137-349B \citep{casewell15} and SDSS1411 \citep{casewell18b} have been shown to be too bright on both the day and night side in the $K$ band when compared to models of unirradiated brown dwarfs, and this may indeed be the same effect.

\subsection{A possible evolutionary history for GD\,1400}

Even if the mass of the white dwarf GD\,1400A is as much as $0.1$ M$_{\odot}$ lower than the spectroscopically determined value of $0.68$ M$_{\odot}$, it is still around the peak of the mass distribution for field white dwarfs \citep{tremblay16} and is most likely a C/O-core degenerate. Thus, the progenitor star likely underwent two giant phases, as would be expected for an isolated field white dwarf. Our progenitor mass as determined by \textsc{wdwarfdate} is  2.09$^{+0.49}_{-0.52}$~M$_{\odot}$ assuming single star evolution.

\begin{figure}
\centering
{\includegraphics[angle=0,scale=0.27,trim={1cm 1.5cm 0 0},clip]{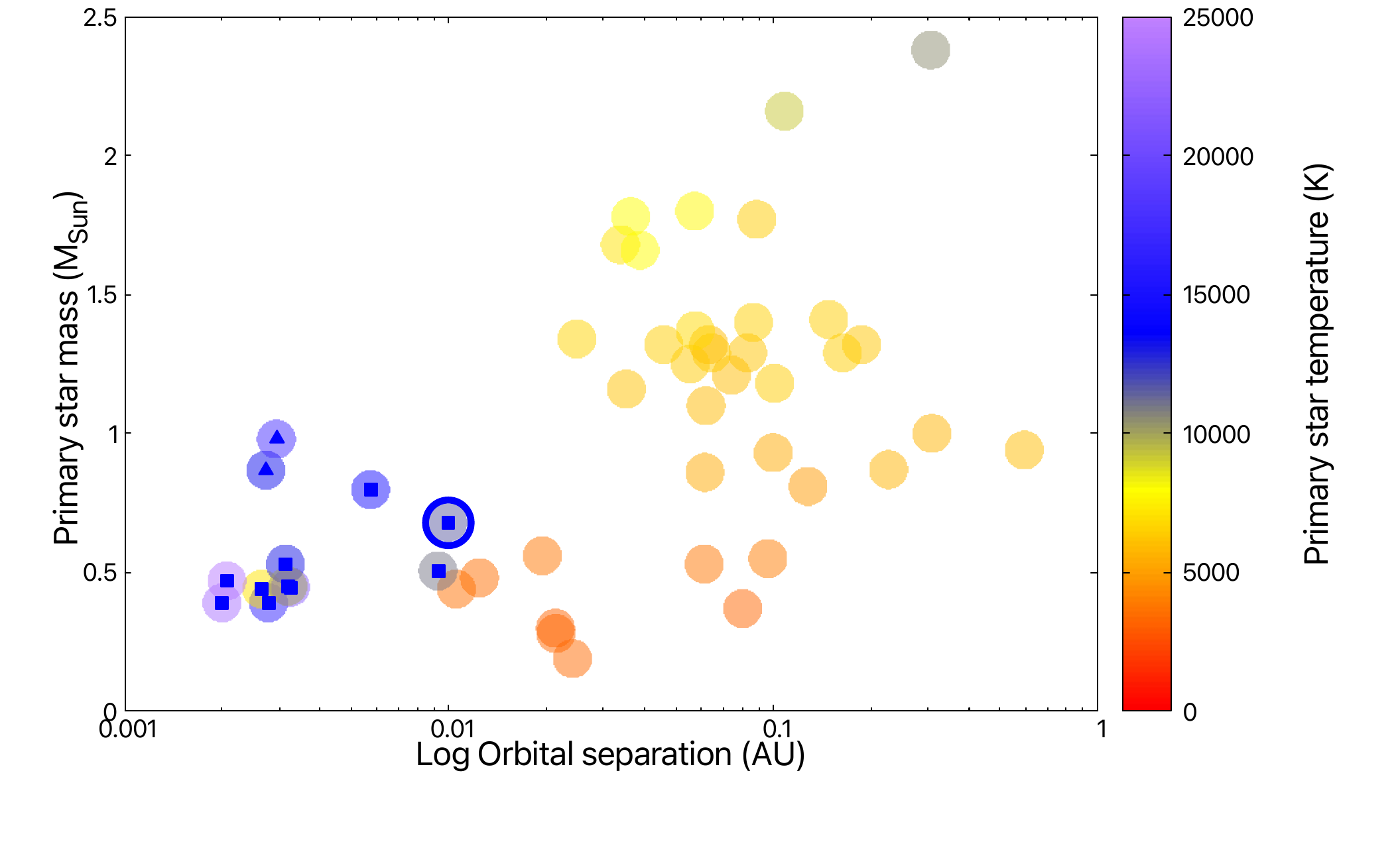}}
\caption{Brown dwarfs occupying the region known as the brown dwarf desert. Transiting brown dwarfs \citep{henderson24} orbiting main sequence stars as well as the known close, detached white dwarf-brown dwarf binaries (filled squares), and cataclysmic variables (filled triangles) are shown comparing the primary masses, effective temperatures and orbital separations. GD\,1400AB is outlined with a circle}\label{fig:mass_sep}
\end{figure}

Assuming solar metallicity, \citet*{Hurley00} estimate the radius of a $2.5$ M$_{\odot}$ RGB star as $25$ R$_{\odot}$ or $0.12$\,AU, and the radius of a $2.5$ M$_{\odot}$ AGB star as $250$ R$_{\odot}$ or $1.2$\,AU. Therefore, during the main sequence phase GD\,1400B must have orbited its parent star at a separation somewhere between $\approx0.1 - 1$\,AU. Interestingly, this region is precisely that which is largely void of brown dwarfs around solar-type main sequence stars \citep{grether06, triaud17}. While only $\sim$40 transiting brown dwarfs have been discovered in this region (e.g. \citealt{henderson24}), coined the brown dwarf desert (Figure \ref{fig:mass_sep}), $\sim$8 per cent are orbiting stars more massive than 2~M$_{\odot}$ \citep{vowell, grieves, psaridi22}, although none of these are giants.
Similarly, RV searches at first ascent giant stars, which are descended from the intermediate-mass stars that are the progenitors of white dwarfs like GD\,1400, find brown dwarfs orbiting within a few AU at roughly the same frequency as for solar-type stars ($<$\,1\%; \citealt{hayes, lovis07, Liu}), and candidate substellar companions to red giants have been identified as causing long secondary period variability \citep{giants}.  

The details of the CE interaction following contact with the expanding AGB envelope are poorly understood, but it is thought the drag on the brown dwarf forces it to quickly spiral in towards the core of the AGB star. The deposition of orbital energy as kinetic energy in the envelope causes it to be ejected from the system, leaving a close binary consisting of the AGB core (now a white dwarf) and the brown dwarf (e.g. \citealt{izzard}). 

In order to test this evolutionary pathway, we have reconstructed the CE phase using the same method recently used by \citet{zorotovic22} for similar systems. This algorithm, developed by \citet{zorotovic10}, searches for possible white dwarf progenitors in a grid of stellar evolution tracks generated with the single-star evolution (SSE) code from \citet{Hurley00}. Assuming that the core mass of the progenitor when it filled the Roche lobe is equal to the mass of the current white dwarf, we used the radius of the progenitor and the companion mass, assuming Roche geometry, to determine the period the system had at the onset of the CE phase for each possible progenitor. We then used the \textit{energy formalism} for CE developed by \citet{webbink84} with the binding and orbital energy calculated as in the binary-star evolution (BSE) code from \citet{hurley02}. The CE efficiency $\alpha_{\rm CE}$ was left as a free parameter, while the structural parameter $\lambda$ was calculated as in \citet{claeys14} without considering contributions from recombination energy. The metallicity was set to z=0.02. For the calculation of the final orbital energy we used the period the system had when it emerged from the CE, which was calculated based on the current orbital configuration and the cooling time of the white dwarf, assuming gravitational radiation during the post CE phase \citep{schreiber+gaensicke03}. 

\begin{figure}
 \centering
  \includegraphics[width=0.49\textwidth,angle=0]{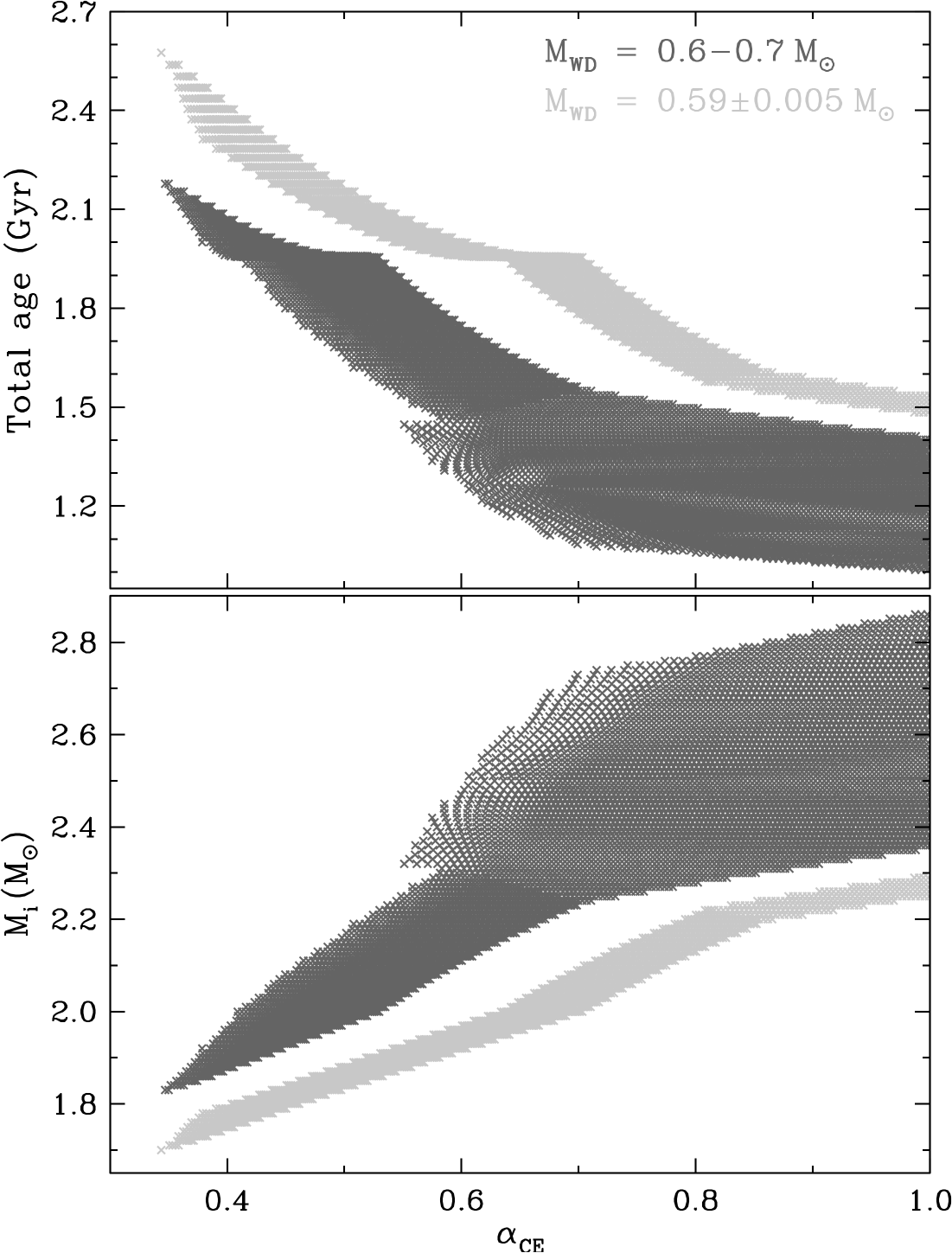}
 \caption{Total age of the system (\textit{top}) and initial mass of the progenitor of the white dwarf (\textit{bottom}) as a function of the CE efficiency $\alpha_{\rm CE}$ derived from our reconstruction. A fixed brown dwarf mass of $0.0812$ M$_{\odot}$ was assumed. For the white dwarf mass, we consider two scenarios: one assuming a possible range of $0.6-0.7$ M$_{\odot}$ (dark gray), and the other adopting $0.590\pm0.005$ M$_{\odot}$ as derived by \citet{walters} (light gray).}
 \label{fig:recons}
 \end{figure}

In Figure\,\ref{fig:recons} we show the total age of the system (top), i.e. the time until the CE phase occurred plus the cooling age of the white dwarf, and the progenitor mass (bottom) as a function of the CE efficiency $\alpha_{\rm CE}$ derived from our reconstruction. The results in dark gray were computed assuming a white dwarf mass consistent with our estimation. Considering the different values reported in the literature for the white dwarf mass, we allowed the progenitor's core to vary between $0.6$ and $0.7$ M$_{\odot}$. The mass of the companion was set to $0.0812$ M$_{\odot}$ based on our estimation derived from the velocity ratio reported by \citet{walters}. 
We were able to reconstruct the CE phase without the need of recombination energy and with a small efficiency (as low as $\alpha_{\rm CE}\sim0.35$), especially if the system is at least 2\,Gyr old. Our reconstruction predicts an initial progenitor mass in the range of $\sim1.83-2.86$ M$_{\odot}$ and a total age of $\sim1-2.2$\,Gyr, with all possible progenitors being on the AGB phase at the onset of the CE evolution. When restricting $\alpha_{\rm CE} \le 0.41$, to align with the range derived by \citet{zorotovic22} for similar systems, the ranges for the possible mass of the progenitor and total age of the system are reduced to $\sim1.83-2.0$ M$_{\odot}$ and $\sim1.95-2.2$\,Gyr, respectively.
While the reconstructed parameters presented in Figure\,\ref{fig:recons} correspond to a fixed metallicity (z=0.02), different metallicities were also tested. Although these results are not shown in the figure, a lower metallicity leads to faster evolution, allowing for possible progenitors with lower masses and less envelope mass to expel during the CE phase. This results in a lower minimum value for the reconstructed efficiency ($\alpha_{\rm CE}$).
 
We also tested our reconstruction using the white dwarf mass of $0.590\pm0.005$ M$_{\odot}$ derived by \citet{walters}. These results are shown in light grey in Figure\,\ref{fig:recons}. Again, only possible progenitors on the AGB phase are found, and the minimum value obtained for $\alpha_{\rm CE}$ remains as low as for a larger white dwarf mass. However, the reconstruction for a lower white dwarf mass predicts progenitors with lower initial masses, resulting in an older system compared to the case with the higher white dwarf mass derived here.

Finally, considering the uncertainty in the brown dwarf mass, we repeated the calculations assuming a mass at the lower end of our estimations, i.e., $0.071$ M$_{\odot}$. While the ranges for the possible initial mass of the progenitor and total age remained almost unchanged, we observed a slight shift towards larger values of the CE efficiency with a minimum value of $\alpha_{\rm CE} \sim 0.4$. This behavior is expected, as lowering the mass of the companion does not affect the possible progenitors of the white dwarf. However, it reduces the available orbital energy to eject the envelope, which translates in a larger CE efficiency required to emerge from the CE phase at a given orbital period.

If the CE phase had happened on the first ascent of the (Red) Giant Branch (RGB) phase, the growth of the progenitor's core would have been significantly truncated. The premature ejection of the envelope during the RGB phase leaves behind a naked He core ($<0.48$ M$_{\odot}$). This core can ignite helium to become a hot subdwarf star, and later evolve into a hybrid He/CO white dwarf if the CE phase occurs near the tip of the RGB \citep{han2002,Arancibia24}. Otherwise, if the naked core is not massive enough to ignite helium, it contracts and cools down after the CE ejection, becoming a He-core white dwarf. This is most likely the evolutionary path for WD\,0137-349A, and the majority of the other known systems.

GD\,1400 appears to have a different evolutionary history to many of the known close, detached white dwarf$+$brown dwarf binaries. Most of them contain white dwarfs with a mass $<0.5$ M$_{\odot}$ \citep[see Table 1 in][]{zorotovic22}.
The WD\,0137-349AB system, for example, had the smaller original separation ($\lesssim$\,0.1\,AU) and underwent CE evolution on the RGB, while GD\,1400B originally orbited its parent star at a wider separation, roughly between $0.1$ and $1$\,AU, and went through CE evolution only when its companion reached the AGB. 

%\begin{figure}
%\centering
%{\includegraphics[angle=0,scale=0.23]{Mass_separation_WD.pdf}}
%\caption{The known close, detached white dwarf-brown dwarf binaries comparing their white dwarf masses and orbital separations. Cataclysmic variables tend to occupy the bottom right of this diagram with higher than average mass white dwarf primaries, whereas Gaia~0007-1605 and WD1856+534Ab are off the top of the figure with orbital separations of 0.17 and 0.02 AU respectively. }\label{fig:mass_sep}

%\end{figure}

\section{Summary}

Radial velocity measurements with UVES on the VLT conclusively demonstrate that the white dwarf $+$ L6-L7 dwarf binary GD\,1400 is a close system with an orbital period $P_{\rm orb} = 9.98$~h. Optical time-series photometry of GD\,1400A shows that the pulsations are consistent with a large-amplitude ZZ~Ceti variable and that the frequencies and amplitudes are unstable from year to year. $ALLWISE$ photometry detects a weak reflection effect suggesting a day-night temperature difference of $\sim$100~K, but a nightside that is warmer than would be suggested by field dwarfs of the same spectral type, suggesting either the system is younger, or the constant irradiation from the white dwarf has slowed the cooling of the brown dwarf. The brown dwarf must have survived a prior phase of CE evolution, when the progenitor of the white dwarf was on the AGB phase and not the RGB phase, making it different to the majority of  the known systems.

\section{Acknowledgments}

%MRB and RN acknowledge the support of STFC Advanced Fellowships. 

SLC acknowledges the support of an STFC Ernest Rutherford Fellowship ST/R003726/1. PB is supported in part by the NSERC Canada and by the Fund FQR-NT (Qu\'ebec). JRF thanks the University of Leicester's College of Science and Engineering for a PhD studentship. We thank Paul Dobbie for his input into the project.
For the purpose of open access, the author has applied a Creative Commons Attribution (CC BY) licence to the Author Accepted Manuscript version arising from this submission.
This paper is based on observations collected at the European Southern Observatory, Paranal, Chile (programmes 077.D-0673(A), 080.C-0587(A)), and observations made at the South African Astronomical Observatory (SAAO). This publication also makes use of data products from the Wide-field Infrared Survey Explorer, which is a joint project of the University of California, Los Angeles, and the Jet Propulsion Laboratory/California Institute of Technology, funded by the National Aeronautics and Space Administration.

%%%%%%%%%%%%%%%%%%%%%%%%%%%%%%%%%%%%%%%%%%%%%%%%%%
\section*{Data Availability}
The UVES and SOFI data are available in the ESO data archive., and the ALLWISE data are public and available at IRSA. The SAAO data are available on request to the authors.

%%%%%%%%%%%%%%%%%%%% REFERENCES %%%%%%%%%%%%%%%%%%

% The best way to enter references is to use BibTeX:

\bibliographystyle{mnras}
\bibliography{refs} % if your bibtex file is called example.bib

% Alternatively you could enter them by hand, like this:
% This method is tedious and prone to error if you have lots of references
%\begin{thebibliography}{99}
%\bibitem[\protect\citeauthoryear{Author}{2012}]{Author2012}
%Author A.~N., 2013, Journal of Improbable Astronomy, 1, 1
%\bibitem[\protect\citeauthoryear{Others}{2013}]{Others2013}
%Others S., 2012, Journal of Interesting Stuff, 17, 198
%\end{thebibliography}

%%%%%%%%%%%%%%%%%%%%%%%%%%%%%%%%%%%%%%%%%%%%%%%%%%

%%%%%%%%%%%%%%%%% APPENDICES %%%%%%%%%%%%%%%%%%%%%

%%%%%%%%%%%%%%%%%%%%%%%%%%%%%%%%%%%%%%%%%%%%%%%%%%

% Don't change these lines
\bsp	% typesetting comment
\label{lastpage}
\end{document}